# Observational Evidence for Wind-Driven Low-Pass Filtering of Infrasound at Short Range

**Elizabeth A. Silber[1], Daniel C. Bowman[1,2,¥], Sasha Egan[3,4], Lawrence Burkett[4], Michael Fleigle[1], Keehoon Kim[5], Tesla Newton[3,4], Loring P. Schaible[1], Richard Sonnenfeld[4], Nora Wynn[1], Jonathan Snively[6]**

[1]Sandia National Labs, Albuquerque, NM, 87123

[2]Pacific Northwest National Laboratory, Richland, WA, 99354

[3]Energetic Materials Research and Testing Center, 1001 South Road, Socorro, NM, 87801

[4]New Mexico Institute of Mining and Technology, Socorro, NM, 87801

[5]Lawrence Livermore National Laboratory, Livermore, CA, 94551

[6]Embry-Riddle Aeronautical University, Daytona Beach, FL, 32114

[¥]now at PNNL

Corresponding author: Elizabeth A. Silber (esilber [at] sandia.gov)

**Key Points:**

- 10-ton ground chemical explosions conducted at the same location in May and October show distinct infrasound wavefields.

- Strong low-altitude winds were found to impact the frequency content of the waves depending on propagation direction.

- Accurate, time-of-event atmospheric specifications are essential for interpreting infrasound signals, even at local (<10s km) range.

**Abstract**

Infrasound from controlled explosions provide a unique opportunity to isolate atmospheric effects on propagation. We report observations from two campaigns in May and October 2024, each featuring 10-ton TNT-equivalent controlled surface chemical explosions recorded by a dense network of 31 single-sensor stations within 23 km. Despite identical sources, the observed wavefields were very different. October signals followed a near-unimodal period–distance trend, whereas May signals exhibited a pronounced azimuthal bifurcation in both period and celerity. Downwind paths largely preserved the short-period baseline observed in October, while upwind paths showed systematically longer periods caused by wind-driven low-pass filtering. This study provides the first direct observational evidence that tropospheric winds can impose azimuth-dependent low-pass filtering at local ranges, without the influence of measured temperature inversions. Thus, the structure of the atmosphere can modify the spectral characteristics of low-frequency acoustic waves even at a distance of only a few kilometers.

**Plain Language Summary**

Scientists use very-low-frequency sound waves, called infrasound, to detect and study large explosions. These waves can travel long distances, but the atmosphere bends and filters them, altering what sensors can record at ground. In May and October 2024 we carried out identical explosive tests (equivalent to 10 tons of TNT) and recorded them with dozens of instruments up to 23 km away. In October, the signals behaved as predicted, following one clear pattern in their timing and length. In May, however, the same blasts produced two different kinds of signals depending on direction: one kept the shorter signal lengths we expected, while the other arrived with longer signal lengths because winds in the atmosphere filtered out the shorter parts of the wave. These results show that the state of the atmosphere at the time of an explosion can strongly affect what is recorded, even at short distances.

## 1 Introduction

Accurate characterization of large explosions remains a central objective of geophysical monitoring, yet the atmosphere through which infrasound propagates is the largest source of uncertainty (e.g., Albert *et al.*, 2023; Averbuch *et al.*, 2022a; Averbuch *et al.*, 2022b; Drob *et al.*, 2003). The Earth's atmosphere functions as a dynamic, stratified acoustic waveguide, capable of supporting low-frequency infrasound propagation over thousands of kilometers (Evans *et al.*, 1972). This property makes infrasound a versatile tool for detecting and characterizing energetic phenomena including volcanic eruptions (e.g., Matoza *et al.*, 2019), bolides (e.g., Pilger *et al.*, 2020; Silber, 2024), lightning (Farges, 2009)) and anthropogenic activities (e.g., re-entry (Clemente *et al.*, 2025; Silber *et al.*, 2025a), mining (Arrowsmith *et al.*, 2008; Ronac Giannone *et al.*, 2025; Stump *et al.*, 2001), and nuclear (Assink *et al.*, 2016) and chemical explosions (Bowman and Krishnamoorthy, 2021)). The efficacy of such applications depends on our ability to separate the source term from atmospheric influences at the receiver.

Sound propagation is governed by the effective sound speed ($c_{eff}$) as a function of altitude ($z$), which is the sum of the adiabatic sound speed, ($c_0$), and the projection of the wind vector ($\vec{v}$) onto the direction of propagation ($\hat{n}$):

$$c_{eff}(z, \hat{n}) = c_0(z) + \vec{v}(z) \cdot \hat{n}. \qquad (1)$$

Here, $c_0(z) = \sqrt{\gamma R T(z)}$, where $T$ is the temperature, $\gamma$ is the ratio of specific heats and $R$ is the specific gas constant. The second term, $\vec{v}(z) \cdot \hat{n}$, is the projection of the wind vector in the direction of propagation. As an infrasound wave travels, it is modified by the complex, stratified structure of the atmosphere (Chunchuzov *et al.*, 2015; Evers and Haak, 2010; Kulichkov, 2000; Negraru and Herrin, 2009; Negraru *et al.*, 2010). Variability in $c_{eff}$ refracts acoustic energy through multiple atmospheric layers, from turbulent boundary layers (Smink *et al.*, 2019) to the seasonally reversing winds of the stratosphere (Butler *et al.*, 2017) and steep thermospheric gradients (Smets and Evers, 2014). The atmosphere thus acts as a planetary-scale acoustic lens, alternately focusing and defocusing sound (Waxler and Assink, 2019), producing ducts that trap energy (Albert *et al.*, 2023; Evans *et al.*, 1972; Sutherland and Bass, 2004) and creating shadow zones that can shift with season (Negraru *et al.*, 2010).

This complex interaction poses a fundamental inversion problem: signals recorded at a distance represent a convolution of the source time function with the atmospheric transfer function (e.g., Green and Nippress, 2019). Reliable estimation of source properties, such as yield, therefore requires disentangling propagation effects, a challenge that is exacerbated by the scarcity of high-resolution atmospheric specifications (e.g., Albert, 2022). Controlled explosions of known yield, timing, and emplacement location provide benchmark datasets for testing propagation models and refining inversion algorithms (Bowman, 2019; Kim *et al.*, 2018; Popenhagen *et al.*, 2023; Silber *et al.*, 2023).

While long-range propagation has received substantial attention (e.g., Arrowsmith *et al.*, 2008), the effects of local environment (< 50 km) remains poorly constrained. It is in this proximal region that the shock-to-acoustic transition occurs, and where the initial interaction with atmospheric structure influences the evolution of the waveforms that are ultimately observed at long distances (e.g., Evers and Haak, 2010). Prior studies have shown that even nominally identical small-yield shots can exhibit strong variability due to mesoscale meteorology (Young *et al.*, 2018), and that minute-scale atmospheric fluctuations can imprint measurable spectral modulations (Dannemann Dugick and Bowman, 2022). Two critical gaps remain: (i) the absence of season-spanning datasets with dense azimuthal coverage at local distances, and (ii) limited opportunities to benchmark modern propagation models against tightly constrained ground truth.

Creating an opportunity to address these gaps, the AtmoSense program's AIRWaveS (Atmosphere-Ionosphere Responses to Wave Signals) project included season-spanning experimental campaigns, designated Bouncing Bilby 1 (BB1) and Bouncing Bilby 2 (BB2), using repeatable, fixed-yield chemical explosions at the New Mexico Institute of Mining and Technology – Energetic Materials Research and Testing Center (EMRTC) in Socorro, New Mexico. Here, we analyze signals from the 10-ton trinitrotoluene (TNT) equivalent (1 ton TNT = $4.184 \cdot 10^9$ J) surface shots conducted in May and October 2024 and recorded by a dense local network extending to 23 km. By comparing these campaigns, we demonstrate that springtime winds imposed a complex, azimuthally structured wavefield absent in autumn. These results provide a rare, local-range benchmark of atmospheric modulation of infrasound, with direct implications for both geophysical monitoring and the development of predictive propagation models.

## 2 Methods

### 2.1 Experimental Campaigns

The two field campaigns  BB1 and BB2, were conducted at EMRTC and here provide an opportunity to investigate local infrasound propagation under different atmospheric conditions. BB1 took place on May 8, 2024, and BB2 on October 16, 2024. Each campaign consisted of five distinct explosive events, comprising four 1-ton and one 10-ton TNT-equivalent chemical charges. The charges were chemical surface detonations, providing a consistent and repeatable source for infrasound generation. This study focuses on the 10-ton events, which provide a consistent, high signal-to-noise baseline for cross-seasonal comparison. The 10-ton shots occurred at 20:29:00 UTC on their respective days (further details are listed in **Table S1**).

To sample the acoustic wavefield, a dense network of single-sensor infrasound stations was deployed around the source region (**Figure 1**). The closest station was located 2.7 km from the detonation and the farthest was 22.3 km. This geometry enabled high-resolution characterization of the transition from shock to an acoustic wave and its initial interaction with the lower atmosphere. The network's broad azimuthal distribution was designed to capture directional propagation effects, particularly those induced by winds. To ensure replicability, a consistent network layout was implemented in both May and October (**Tables S2 and S3**). A deliberate design feature was the grouping of stations into two approximately orthogonal azimuthal clusters: the WCL line oriented at ~253° and the PPL line (including station SLVR) oriented at ~22° relative to geographic north (**Figure 1**), providing a configuration optimized for testing propagation anisotropy.

All stations were equipped with Gem infrasound microbarometers: portable, low-power sensors developed for rapid deployment in large-scale geophysical campaigns. Operating at a 100

Hz sampling rate and a ± 100 Pa dynamic range, the Gems provide digital waveforms well-suited for detailed analysis of acoustic signals originating from both ground-based (Anderson *et al.*, 2018) and airborne impulsive sources (Silber *et al.*, 2024). The use of the same instruments across the network minimized bias and ensured that observed signal variations reflected propagation effects rather than differences in sensor response (Silber and Bowman, 2025).

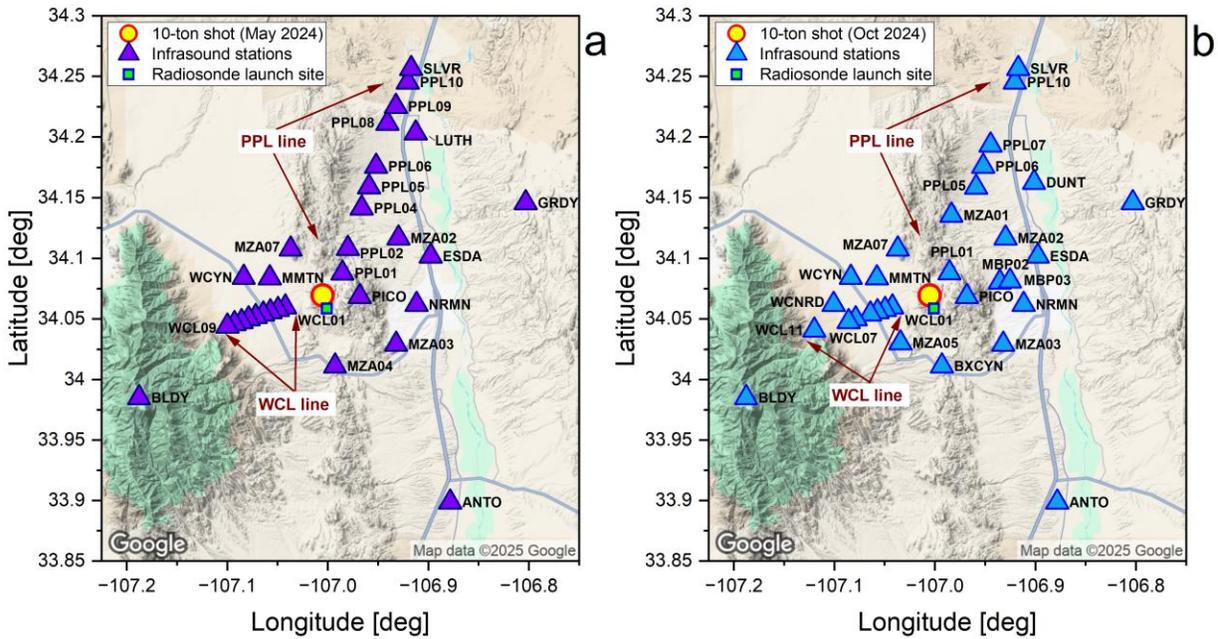

**Figure 1.** Deployment map of the infrasound sensor network around EMRTC during the May (a) and October (b) campaigns. The chemical explosion source is marked by the red-yellow circle, and the locations of the single-sensor stations are shown as triangles. The radiosonde launch site is indicated with the square.

### 2.2 Infrasound Data and Processing

The recorded time-series data were processed to extract relevant signal parameters using a consistent methodology (Silber and Bowman, 2025). Raw pressure waveforms were first filtered with a fourth-order, zero-phase Butterworth filter to isolate infrasound signals from background noise. To evaluate sensitivity to filter parameters, each record was processed across multiple bands with lower frequency cutoffs between 0.1 and 0.7 Hz and upper frequency cutoffs between 10 and 45 Hz. Similar estimates were derived using a 0.1 Hz high-pass filter.

From these filtered waveforms, we measured the maximum peak-to-peak (P2P) amplitude across the principal lobe of the N-wave, defined between the first zero crossings bracketing the maximum positive pressure excursion. Because near-source ground coupling, site heterogeneity, and sensor dynamic range can confound absolute amplitude, these values are reported for completeness in the Supplement (**Tables S2, S3, Figure S2**) but not interpreted further. Our main analysis emphasizes signal celerity and period, which are less sensitive to local coupling.

The signal period was determined using zero crossings at the point of maximum amplitude (Ens *et al.*, 2012; Revelle, 1997). For the characteristic N-wave shape, the positive compression phase is the most stable and least affected by scattering or coda contamination. Following Silber and Bowman (2025), we measured the duration of this half-cycle and doubled it to estimate the full-cycle period. Uncertainties in both period and amplitude are reported as interquartile ranges across the analysis bands, with full measurement sets given in the Supplement.

To diagnose the propagation paths, we examined the signal celerity, the apparent horizontal propagation speed defined as $v_{app} = r/\Delta t$, where $r$ is the geodesic source–receiver distance and $\Delta t$ is the travel time relative to detonation. Here, celerity represents the path-integrated apparent propagation velocity inferred from travel times, while effective sound speed, discussed in later sections, describes the local wind-modified phase speed derived from radiosonde profiles. Different propagation paths have distinct geometric lengths and thus yield characteristic celerities (Negraru and Herrin, 2009; Negraru *et al.*, 2010). To characterize the prevailing atmospheric state, three radiosonde soundings were launched during each campaign. The profiles verified the absence of strong low-level temperature inversions and provided vertical profiles of the wind, adiabatic sound speed ($c_0$), and effective sound speed ($c_{eff}$) to altitudes of 16-17 km, which form the basis for interpreting observed propagation differences. Effective sound-speed profiles were computed from the radiosonde measurements using Eq. (1), projecting the measured wind vectors onto the azimuths of interest and assuming a solely horizontally stratified atmosphere

## 3 Results and Discussion

### 3.1 Seasonal Observations of the Wavefield

The results from the two campaigns provide observational evidence that low-altitude atmospheric variability profoundly modulates the local infrasound wavefield. **Figure 2**, which summarizes signal period (**a,c**) and celerity (**b.d**) versus distance for both seasons, reveals a notable dichotomy between May and October experiments.

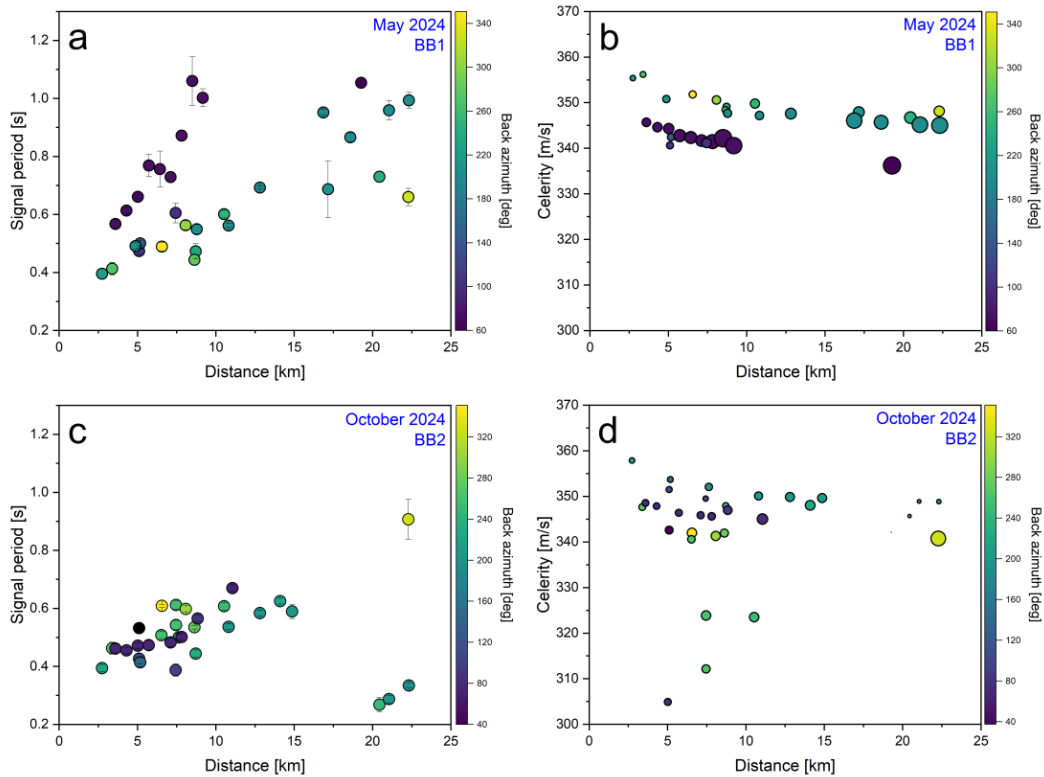

**Figure 2.** Signal period (left) and celerity (right) versus distance for the May (top panels a, b) and October (bottom panels c, d) chemical explosions. May shows clear azimuthal splitting; October is near-unimodal. Because stations are single sensors (no array processing), back azimuth shown is purely geometric. All data points in panels (a,c) include error bars representing the interquartile range across analysis frequency bands; for most points these are smaller than the symbols and therefore not visible.

The October experiment serves as an isotropic baseline, revealing what appears to be a single ensemble of signal arrivals regardless of azimuth. Signal periods follow a coherent trend with a single (-unimodal) population (**Figure 2c**), but the lack of stations 15–20 km means that continuity across this range is inferred rather than directly observed. Celerities exhibit only moderate scatter (**Figure 2d**). This variability is consistent with propagation through a relatively quiescent lower atmosphere, where small-scale turbulence and weak refractive inhomogeneities introduce travel-time jitter but do not impose systematic directional dependence (Attenborough, 2002; Ostashev and Wilson, 1997; Wilson, 1996). In contrast, the May data exhibit pronounced azimuthal bifurcation in both period and celerity (**Figure 2a,b**). Two distinct ensembles of arrivals are evident: one characterized by faster apparent celerities and shorter periods, the other by slower celerities and progressively longer periods with range. This azimuthally organized splitting, absent in October, indicates that the atmosphere imposed a strong anisotropy on the local acoustic wavefield.

### 3.2 Synoptic-Scale Cause of the Anisotropy

The atmospheric origin of this contrast is evident in the radiosonde-derived effective sound speed profiles (**Figure 3**). In May 2024, $c_{eff}/c_0$ departs markedly from unity and exhibits strong azimuthal dependence. In contrast, the October 2024 profiles are less distinguishable between azimuths, indicating more isotropic conditions. This contrast highlights that in May the atmospheric winds were sufficient to impose azimuthal asymmetries in wave propagation, where in October they were not. Other mechanisms, such as topographic shielding or ground reflections, contribute to waveform complexity but cannot explain the clear seasonal dichotomy, as these factors remained constant between campaigns.

Shallow structures capable of ducting acoustic energy have previously been associated with strong temperature inversions, such as those observed following the Buncefield explosion (Green *et al.*, 2008). Our study, however, provides evidence for a purely wind-driven mechanism at short ranges. Whereas previous research has largely demonstrated such effects for long-range, stratospherically ducted propagation (e.g., Ceranna *et al.*, 2009), the present results establish direct observational evidence that local winds alone can generate azimuthally distinct ducts and act as signal filters at ranges of only a few kilometers. Although there was evidence of low altitude inversions in the radiosonde data (**Figure S3**), this temperature effect alone is not expected to produce a directional effect in the frequency/distance relationship because the sound is affected the same way regardless of azimuth. This finding emphasizes that atmospheric state specifications are not only critical for interpreting far-field infrasound but are equally indispensable for understanding local propagation, where the tropospheric wind field impacts both waveform character and spectral content.

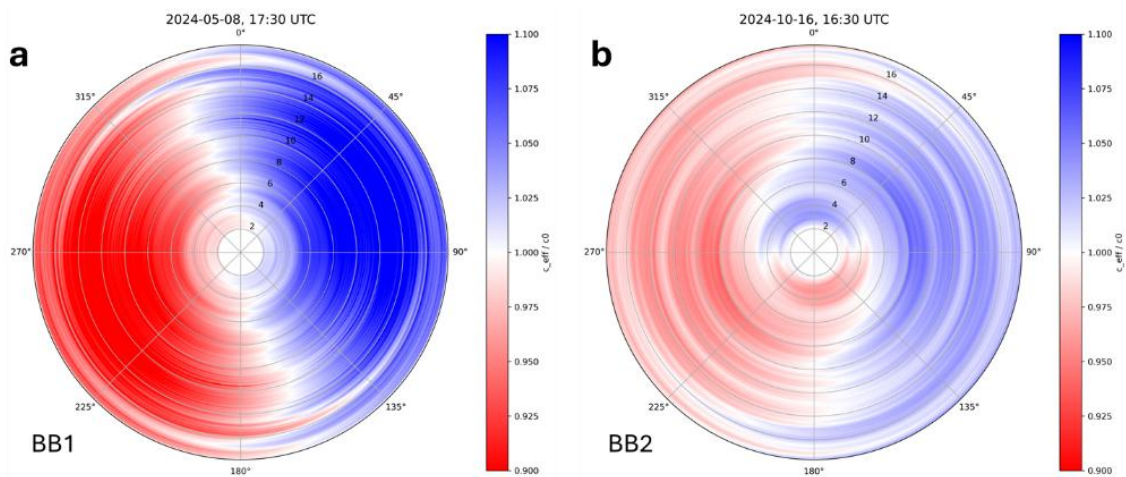

**Figure 3.** Radiosonde-derived profiles of the ratio between effective and adiabatic sound speed ($c_{eff}/c_0$) as a function of altitude for (a) BB1 and (b) BB2. Radial axis denotes altitude in km. The May profile exhibits a strong, vertically coherent easterly jet, producing pronounced anisotropy in effective sound speed between azimuths, whereas the October profile reflects weak and disorganized winds consistent with more-isotropic conditions.

### 3.3 Propagation Mechanisms and Wind-Induced Filtering

To examine the physical mechanism of the anisotropy, we consider the two azimuthal sensor groups: ~22° and ~253°. This configuration was explicitly designed to test for anisotropy **(Figure 1)**. The dynamical origin of this anisotropy is evident in the projected wind velocities and effective speed of sound (**Figure 4**). In May, a coherent easterly jet exceeding 40 m/s, as seen directly from the campaign radiosonde (Figure 4), produced strong tailwinds along 22° and opposing headwinds along 253°. These wind projections map directly onto the sound speed asymmetries in **Figure 3**.

The tailwind environment increases $c_{eff}$, producing a shallow tropospheric duct that favors efficient transmission, higher apparent celerities, and shorter observed periods (Green *et al.*, 2008). The headwind environment reduces $c_{eff}$, forcing upward refraction of energy. Upward-refracted rays undergo frequency-dependent attenuation, preferentially removing high-frequency content (e.g., Kim and Rodgers, 2017; Le Pichon, 2015; Wilson *et al.*, 2015). At these local ranges (≤23 km), the refraction described here refers to continuous upward bending of the acoustic path under headwind conditions, which modifies spectral content; it does not produce a discrete secondary arrival at the sensors. Previous studies have highlighted the role of strong temperature gradients in producing similar ducting and filtering effects and noted that winds can provide an additive effect (Ceranna *et al.*, 2009; Green *et al.*, 2008). Here, however, we provide evidence that winds alone, specifically the strong headwind structures present in the May atmosphere, are sufficient to drive the observed low-pass filtering. This filtering explains why longer periods and slower celerities dominate the upwind arrivals recorded along the WCL azimuth in May. In October, projected winds were weaker, producing similar $c_{eff}$ profiles across azimuths. In this more isotropic atmosphere, signals propagated with little directional dependence, consistent with the lack of any significant differences in recorded celerity or period. Based upon our observations, the observed dichotomy in infrasonic parameters (**Figure 2**) in May is therefore a direct and quantifiable manifestation of changes in tropospheric wind structure.

Here, however, we show that this upward refraction is driven solely by the strong headwind structures present in May, directly linking the observed anisotropy to low altitude winds. While this mechanism explains the systematic azimuthal bifurcation, it does not account for the scatter seen within each group. These additional features point to finer-scale processes, most notably turbulence (see **Section 3.4**).

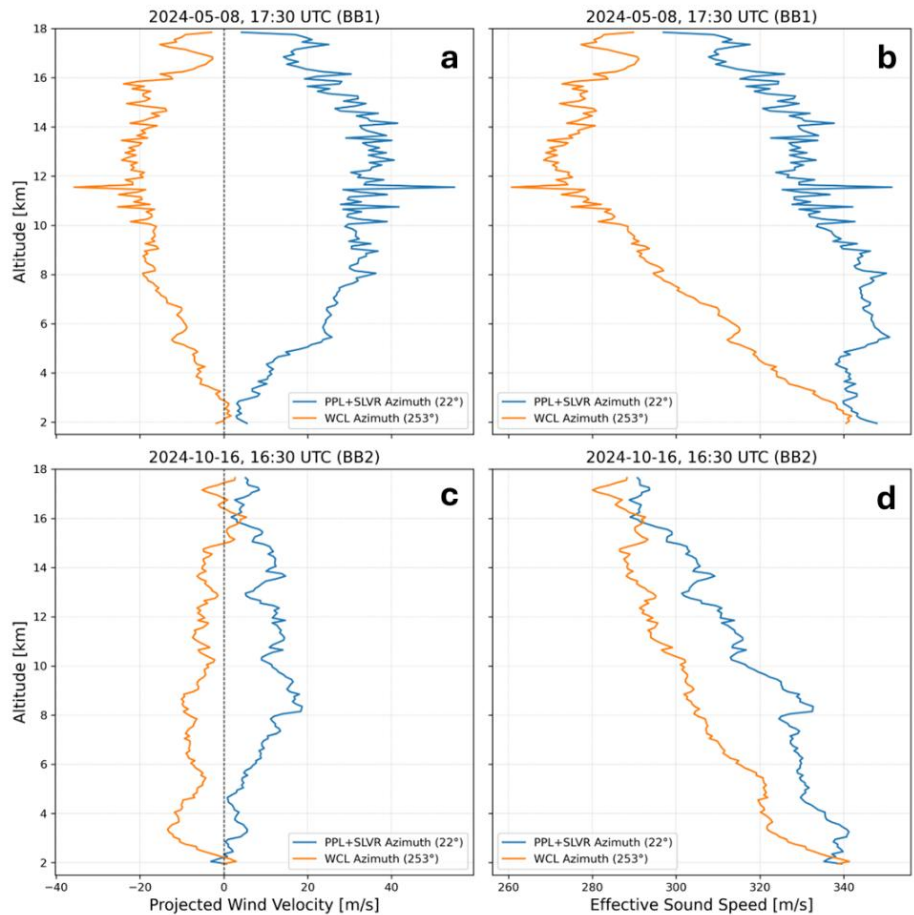

**Figure 4.** Projected wind velocity and $c_{eff}$ profiles along the 22° (blue) and 253° (orange) azimuths from radiosonde soundings on May 8, 2024 ( a-b) and October 16, 2024 (c-d). Panels (a,c) show projected wind velocities, and panels (b,d) show the corresponding effective sound speed profiles. In May, a vertically coherent easterly jet exceeding 40 m/s produced strong tailwinds along ~22° and opposing headwinds along ~253°. In contrast, the October profiles show weak, nearly symmetric winds and minimal azimuthal differences in effective sound speed, indicative of near-isotropic conditions.

### 3.4 Role of Turbulence

The residual scatter evident in both celerity and period requires additional explanation beyond the large-scale ducting and refraction mechanism (Section 3.3). While headwind-driven upward refraction accounts for the systematic azimuthal bifurcation, the variability within each azimuthal group points to additional processes. Radiosonde profiles cannot resolve the temporal evolution of fine-scale anisotropic structures of the lower troposphere, such as thin shear layers, gravity-wave disturbances, and mountain-induced flows (Ostashev and Wilson, 1997; Sabatini *et al.*, 2019), which are likely contributors to the scatter evident even in the otherwise homogeneous October dataset. The distinction between refraction and turbulence in this analysis is based on their spatial signatures: refraction produces systematic, azimuthally coherent differences in period and celerity, whereas we speculate that turbulence manifests as stochastic scatter within each azimuth group.

A useful diagnostic is the contrast between October and May. The October records define the intrinsic, source-controlled baseline under minimal distortion. Downwind PPL+SLVR records in May largely preserved these short periods, while upwind WCL records exhibited systematically longer periods. This discrepancy indicates that the observed bimodality is not anomalous but reflects the combined effects of headwind refraction and turbulence (e.g., Wilson *et al.*, 2015).

The propagation of a finite-amplitude N-wave, such as that generated by an explosion, is fundamentally affected by turbulence. As the shock propagates, fluctuations in wind and temperature fields scatter acoustic energy, producing local perturbations in phase and amplitude that distort the waveform. This process increases the apparent period of the signal, a phenomenon well established in acoustic theory (Pierce, 2019). In the October campaign, we speculate that weak winds generated only modest turbulence, such that broadening was limited to boundary-layer processes and the observed periods remained short. In May, however, the strong and vertically sheared tropospheric jet likely produced a highly anisotropic turbulent environment. Interaction of the N-wave with this turbulent field may have increased scattering, resulting in systematic waveform broadening and period 'lengthening' across all azimuths relative to the October baseline.

We explain the divergence between downwind and upwind signals in May as the superposition of two processes. Turbulence-induced broadening acted on all propagation paths, while the upwind WCL path was additionally subject to refractive filtering. The strong headwind reduced the effective sound speed and forced upward refraction (Kim and Rodgers, 2017). Upward refracted rays experience frequency-dependent attenuation, with high-frequency energy scattered or absorbed more efficiently than low-frequency components (e.g., Le Pichon, 2015; Wilson *et al.*, 2015). This natural low-pass filtering mechanism selectively removes short-period content from the waveform, compounding the turbulent broadening and yielding the longest observed periods in the dataset. The combination of these mechanisms provides the physical explanation for the statistically significant period splitting between azimuths (**Section 3.2**).

These results demonstrate that infrasound propagation and receptions of signals at ground are sensitive to the combined influence of background wind structure and likely the effects of turbulence, as inferred here, even at distances of less than 15 km. The October campaign establishes the source-controlled baseline, while the May campaign reveals how anisotropic wind fields and likely increased turbulence systematically reshape the waveform. The recorded N-waveforms are consistent with signals that have passed through the shock-to-acoustic transition and now propagate in the weak-shock to linear regime, where their shape is preserved but further modified by atmospheric structure. For a 10-ton surface explosion, the shock front decays to a weak-shock or acoustic wave within ~1 km of the source, well inside the instrumented range (Brode, 1955; Kim *et al.*, 2021; Kinney and Graham, 1985; Schnurr *et al.*, 2020). By providing direct observational evidence of turbulence–refraction coupling at short range, this study illustrates the necessity of accurate atmospheric characterization for infrasound monitoring. The findings confirm that even local propagation cannot be interpreted without accounting for state-dependent filtering effects, and they highlight the risk of systematic bias in event characterization if such mechanisms are neglected. Furthermore, they support a future pathway for modeling, to test hypotheses on the local environment effects on infrasound propagation – both on paths and signal evolutions.

### 3.5 Implications

Our results demonstrate that dense, multi-azimuthal networks are essential for resolving atmospheric propagation effects that cannot be inferred from sparse observations or climatological models. Only with broad azimuthal coverage was it possible to detect the systematic bifurcation in celerity and period and to separate source characteristics from path-dependent variability.

Signal period, often regarded as a robust proxy for yield (Revelle, 1997), originates from empirical relations first developed for stationary nuclear explosions and later adapted for other sources as well. However, it is not universally reliable in the presence of variable atmospheric structure. In October, periods reflected the intrinsic source-controlled baseline. Under strong anisotropy in May, however, upwind filtering suppressed high-frequency energy and left longer periods dominant, which, if uncorrected, would bias yield estimates upward. This conclusion is consistent with recent studies of bolide infrasound, which have demonstrated that period–energy relations exhibit significant state dependence and multi-parameter sensitivity (Silber *et al.*, 2025b). Our dataset extends this to controlled explosions, confirming that even under repeatable source conditions, atmospheric filtering can introduce variability large enough to complicate inversion. A climatological analysis using global reanalysis data (e.g., ERA5) to quantify the frequency and persistence of such wind-driven anisotropy would be a natural extension of this work and should be addressed in future studies.

## 4 Conclusions

Controlled 10-ton surface chemical explosions in contrasting seasonal environments provided direct observational evidence that, even at local ranges, infrasonic propagation is strongly influenced by the interaction of synoptic-scale winds and fine-scale turbulence. In May, a coherent easterly jet imposed marked anisotropy in received signals, producing faster, ducted arrivals downwind and slower, low-pass-filtered arrivals upwind, while the October atmosphere yielded near-isotropic propagation. Statistical tests confirm that these effects are significant within 15 km, establishing that signal period, often considered a robust metric, is state dependent and must be interpreted with caution. These results constitute direct observational evidence that tropospheric winds can produce azimuth-dependent low-pass filtering at local ranges, providing new insight into how atmospheric structure affects infrasound propagation. The dense single-sensor dataset assembled here provides a benchmark for validating predictive models and highlights the necessity of incorporating time-of-event atmospheric specifications into monitoring practice.

SAND2026-16694O

## Acknowledgments

We sincerely thank David Green (AWE Blacknest) for useful discussions that helped interpret our results. We appreciate permissions granted by the Bureau of Land Management and the U. S. Forest service, as well as numerous private landowners, for the deployment and operation of the acoustic sensors. This article has been authored by an employee of National Technology & Engineering Solutions of Sandia, LLC under Contract No. DE-NA0003525 with the U.S. Department of Energy (DOE). The employee owns all right, title and interest in and to the article and is solely responsible for its contents. The United States Government retains and the publisher, by accepting the article for publication, acknowledges that the United States Government retains a non-exclusive, paid-up, irrevocable, world-wide license to publish or reproduce the published form of this article or allow others to do so, for United States Government purposes. The DOE will provide public access to these results of federally sponsored research in accordance with the DOE Public Access Plan https://www.energy.gov/downloads/doe-public-access-plan. PNNL is operated by Battelle for the U. S. Department of Energy under contract DE-AC05-76RL01830. This paper describes objective technical results and analysis. Any subjective views or opinions that might be expressed in the paper do not necessarily represent the views of the U.S. Department of Energy or the United States Government, and no official endorsement should be inferred.

## Funding

The AIRWaveS Project and "Bouncing Bilby" experiments were supported under the DARPA Defense Sciences Office (DSO) AtmoSense program via a cooperative agreement HR00112320043 to Embry-Riddle Aeronautical University and agreement O2502097089086567 to Sandia National Laboratories, through which New Mexico Tech's EMRTC was supported under a subcontract. This work is approved for public release; distribution is unlimited.

## Open Research

Radiosonde data collected during each campaign and the waveform data collected from 10-ton shots is available at Zenodo, doi: 10.5281/zenodo.17247825.

## Conflict of Interest Disclosure

The authors declare there are no conflicts of interest for this manuscript.

SAND2026-16694O

Supporting Information for

**Observational Evidence for Wind-Driven Low-Pass Filtering of Infrasound at Short Range**

Elizabeth A. Silber[1], Daniel C. Bowman[1,2,¥], Sasha Egan[3,4], Lawrence Burkett[4], Michael Fleigle[1], Keehoon Kim[5], Tesla Newton[3,4], Loring P. Schaible[1], Richard Sonnenfeld[4], Nora Wynn[1], Jonathan Snively[6]

[1]Sandia National Labs, Albuquerque, NM, 87123; [2]Pacific Northwest National Laboratory, Richland, WA, 99354; [3]Energetic Materials Research and Testing Center, 100 East Road, Socorro, NM, 87801; [4]New Mexico Institute of Mining and Technology, Socorro, NM, 87801; [5]Lawrence Livermore National Laboratory, Livermore, CA, 94551; [6]Embry-Riddle Aeronautical University, Daytona Beach, FL, 32114; [¥]now at PNNL

**Contents of this file**

Figures S1 to S3
Tables S1 to S5
Supplemental Text S1

**Introduction**

These Supplementary Materials include Figures S1–S3, Tables S1–S5, and Section S1. Figures S1–S2 show representative infrasound waveforms and amplitude–distance relationships for the 10-ton chemical explosions conducted in May and October 2024, while Figure S3 presents additional supporting data. Tables S1–S3 list the timing and locations of the explosions and all station-level measurements for both campaigns. Section S1 contains explanatory text and the statistical analysis results summarized in Tables S4 and S5.

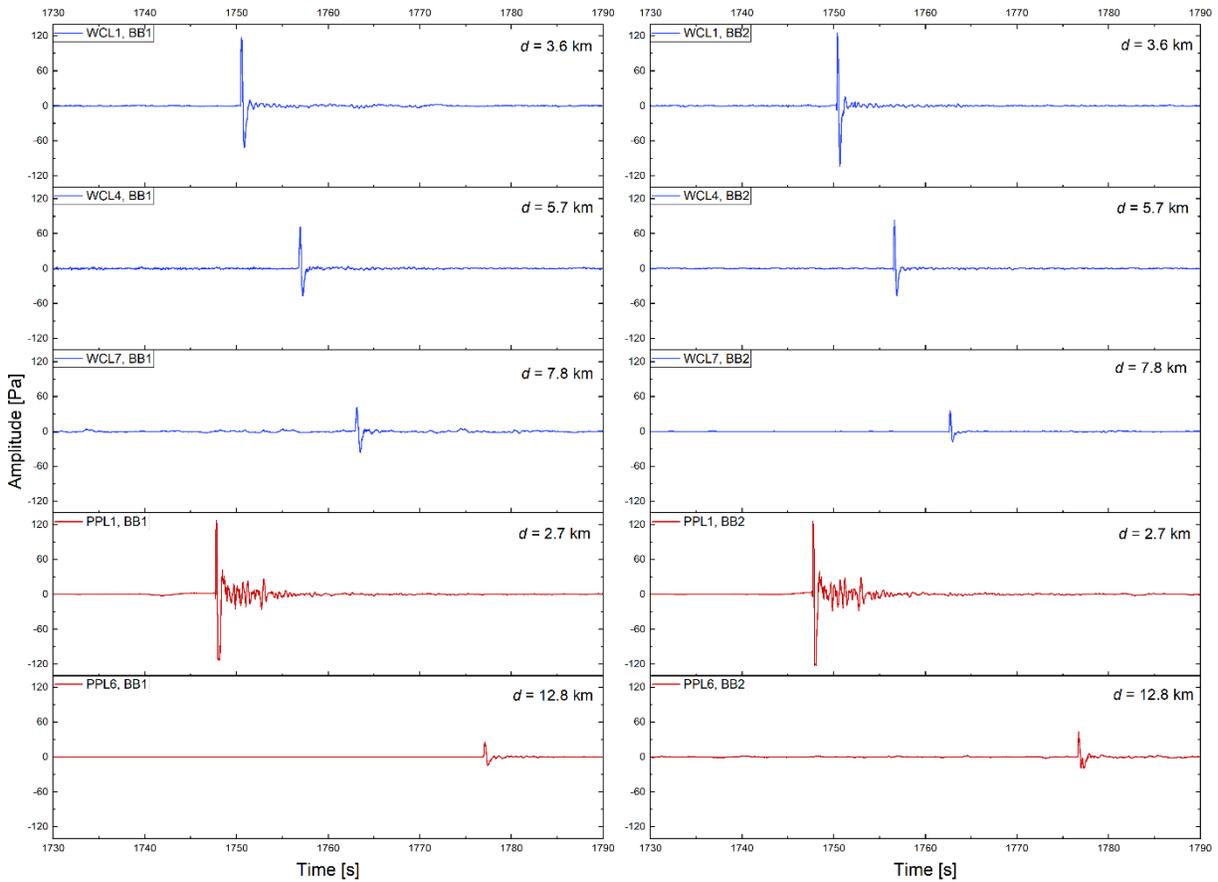

**Figure S1:** Example filtered infrasound waveforms from the 10-ton explosions recorded along the WCL (blue) and PPL (red) station lines during the May 2024 (BB1, left) and October 2024 (BB2, right) campaigns. Signals were band-pass filtered between 0.1 and 20 Hz. Each panel lists the station distance ($d$) from the source, and all panels share a vertical amplitude range of −140 to 140 Pa for direct comparison. The traces show single, high signal-to-noise (SNR) N-wave arrivals with amplitude decay as distance increases. Several near-source records exhibit partial amplitude clipping, making absolute amplitudes unreliable but preserving waveform shape for analysis.

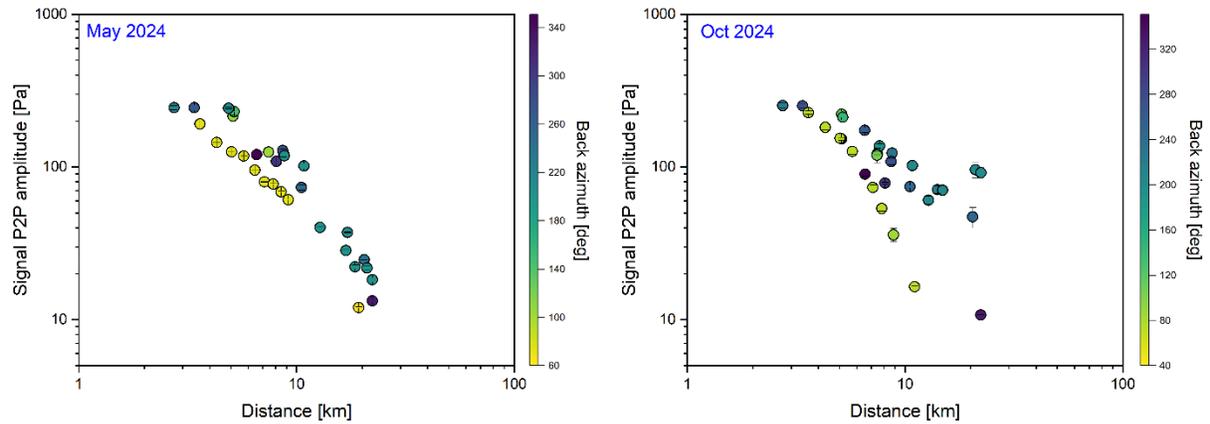

**Figure S2:** Peak-to-peak (P2P) signal amplitude versus distance for the 10-ton explosions in (left) May 2024 and (right) October 2024. Symbols are colored by the geometric back azimuth to the source. Error bars denote interquartile variability from multi-band filtering. Both campaigns show systematic amplitude decay with distance, broadly consistent with spherical spreading and atmospheric absorption. Scatter at fixed range reflects local site effects, coupling differences, and azimuthal propagation variability. Some near-source records exhibit partial amplitude clipping, making absolute levels unreliable. Although amplitudes were measured for all stations, they are not used in the main analysis because they are strongly affected by near-source coupling, sensor siting, and local heterogeneity. The amplitude patterns are provided here for completeness, illustrating the overall decay trend and the azimuthal scatter that accompanies the more robust period and celerity observations discussed in the main text.

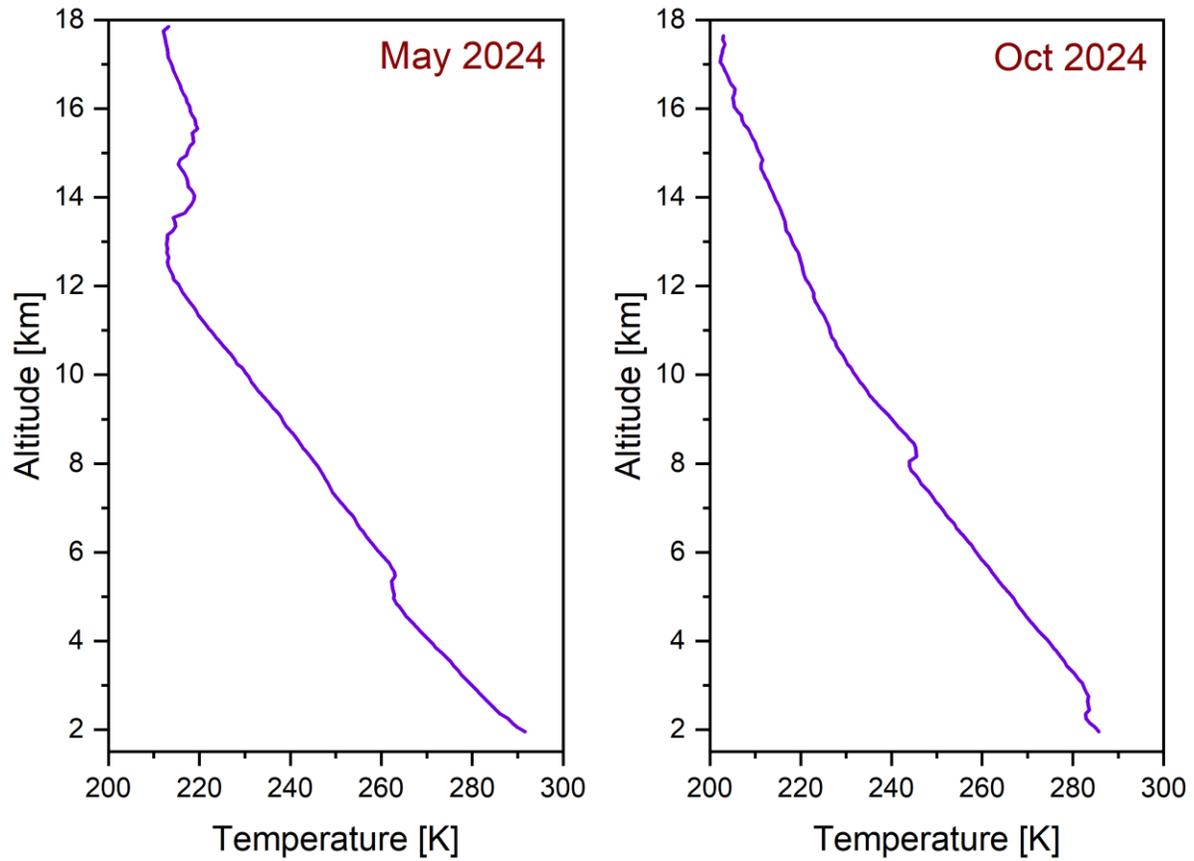

**Figure S3:** Radiosonde-derived temperature profiles for May 2024 (left) and October 2024 (right), illustrating the vertical thermal structure up to 18 km altitude. The soundings were obtained near 17:30 UTC and 16:30 UTC, respectively.

**Table S1:** Chemical explosion experiment particulars. The position uncertainty is ~5 m.

|  | BB1 (May 8, 2024) | BB2 (Oct 16, 2024) |
|---|---|---|
| Time [UTC] | 20:29:00.0000074 | 20:28:59.9999854 |
| Latitude [deg] | 34.0693006128784 | 34.06930918 |
| Longitude [deg] | -107.004971526782 | -107.0050267 |
| Chemical charge [kg] | 9,069.00 ± 2.40 | 9,071.39 ± 2.40 |

**Table S2:** Station-level measurements for the May 8, 2024 (BB1) 10-ton chemical surface explosion. Columns: station ID; latitude, longitude; geodesic range (km) from 34.0693° N, 107.00497° W; geometric azimuth and back azimuth (degrees from north); signal travel time relative to the detonation; apparent celerity; dominant signal period (s) with uncertainty; peak-to-peak amplitude (Pa) with uncertainty. Signal period and amplitude are means over all band sets (see Section 2.2 in the main text). All stations are single sensors; azimuths are geometric, not direction-of-arrival estimates.

| BB1 Station code | Latitude | Longitude | Distance [km] | Azimuth [deg] | Back azimuth [deg] | Observed travel time [s] | Celerity [m/s] | Dominant signal period | Error | Peak-to-peak amplitude [Pa] | Error |
|---|---|---|---|---|---|---|---|---|---|---|---|
| ANTO | 33.8987 | -106.8781 | 22.3 | 148.3 | 328.4 | 64.02 | 348.12 | 0.660 | 0.030 | 13.28 | 0.27 |
| BLDY | 33.9850 | -107.1877 | 19.3 | 240.9 | 60.8 | 57.33 | 336.2 | 1.054 | 0.017 | 12.01 | 0.13 |
| ESDA | 34.1017 | -106.8975 | 10.5 | 70.0 | 250.0 | 30.11 | 349.8 | 0.601 | 0.013 | 73.48 | 2.95 |
| GRDY | 34.1457 | -106.803 | 20.4 | 65.4 | 245.5 | 58.96 | 346.7 | 0.730 | 0.018 | 24.70 | 0.11 |
| LUTH | 34.2033 | -106.9125 | 17.2 | 29.7 | 209.8 | 49.33 | 347.9 | 0.687 | 0.098 | 37.31 | 0.25 |
| MMTN | 34.0839 | -107.0576 | 5.1 | 288.5 | 108.5 | 15.00 | 340.6 | 0.474 | 0.011 | 214.77 | 3.00 |
| MZA02 | 34.1167 | -106.9296 | 8.7 | 52.7 | 232.8 | 24.97 | 349.1 | 0.473 | 0.027 | 123.36 | 7.07 |
| MZA03 | 34.0291 | -106.932 | 8.1 | 123.6 | 303.7 | 23.04 | 350.6 | 0.563 | 0.015 | 108.82 | 0.48 |
| MZA04 | 34.0112 | -106.9928 | 6.6 | 170.2 | 350.2 | 18.65 | 351.8 | 0.489 | 0.014 | 120.53 | 6.10 |
| MZA07 | 34.1076 | -107.0369 | 5.2 | 325.4 | 145.4 | 15.12 | 342.4 | 0.500 | 0.008 | 230.23 | 1.35 |
| NRMN | 34.0620 | -106.9116 | 8.6 | 95.4 | 275.4 | 24.79 | 348.4 | 0.443 | 0.018 | 128.18 | 3.14 |
| PICO | 34.0685 | -106.9682 | 3.4 | 91.4 | 271.5 | 9.52 | 356.2 | 0.413 | 0.022 | 244.60 | 9.44 |
| PPL01 | 34.0879 | -106.9855 | 2.7 | 41.0 | 221.0 | 7.71 | 355.4 | 0.395 | 0.017 | 244.37 | 8.07 |
| PPL02 | 34.1080 | -106.9799 | 4.9 | 28.2 | 208.2 | 13.92 | 350.8 | 0.491 | 0.011 | 242.79 | 2.88 |
| PPL04 | 34.1415 | -106.9662 | 8.8 | 24.0 | 204.0 | 25.27 | 347.6 | 0.549 | 0.016 | 118.37 | 1.50 |
| PPL05 | 34.1588 | -106.9588 | 10.8 | 23.1 | 203.1 | 31.16 | 347.2 | 0.562 | 0.007 | 101.58 | 3.96 |
| PPL06 | 34.1759 | -106.9519 | 12.8 | 22.4 | 202.4 | 36.88 | 347.6 | 0.692 | 0.006 | 40.21 | 0.32 |
| PPL08 | 34.2113 | -106.9406 | 16.9 | 20.5 | 200.6 | 48.73 | 346.1 | 0.951 | 0.010 | 28.40 | 0.40 |
| PPL09 | 34.2251 | -106.9321 | 18.6 | 21.1 | 201.2 | 53.74 | 345.7 | 0.866 | 0.015 | 22.17 | 0.67 |
| PPL10 | 34.2452 | -106.9204 | 21.1 | 21.7 | 201.7 | 60.99 | 345.2 | 0.959 | 0.034 | 21.77 | 0.77 |
| SLVR | 34.2563 | -106.9169 | 22.3 | 21.3 | 201.3 | 64.68 | 345.0 | 0.993 | 0.028 | 18.25 | 0.61 |
| WCL01 | 34.0598 | -107.0423 | 3.6 | 253.0 | 72.9 | 10.42 | 345.7 | 0.567 | 0.013 | 191.46 | 1.22 |
| WCL02 | 34.0579 | -107.0496 | 4.3 | 252.8 | 72.8 | 12.48 | 344.6 | 0.613 | 0.010 | 144.92 | 1.01 |
| WCL03 | 34.0559 | -107.0571 | 5.0 | 252.7 | 72.7 | 14.61 | 344.3 | 0.660 | 0.006 | 125.57 | 0.56 |
| WCL04 | 34.0540 | -107.0643 | 5.7 | 252.7 | 72.6 | 16.70 | 342.7 | 0.768 | 0.038 | 118.18 | 0.66 |
| WCL05 | 34.0519 | -107.0716 | 6.4 | 252.5 | 72.5 | 18.81 | 342.3 | 0.756 | 0.062 | 95.45 | 0.39 |
| WCL06 | 34.0501 | -107.0787 | 7.1 | 252.6 | 72.6 | 20.84 | 341.7 | 0.729 | 0.017 | 79.93 | 0.35 |
| WCL07 | 34.0477 | -107.0857 | 7.8 | 252.2 | 72.1 | 22.89 | 341.5 | 0.872 | 0.018 | 77.41 | 0.64 |
| WCL08 | 34.0460 | -107.0928 | 8.5 | 252.3 | 72.2 | 24.83 | 342.2 | 1.060 | 0.084 | 68.86 | 0.57 |
| WCL09 | 34.0441 | -107.0998 | 9.2 | 252.2 | 72.2 | 26.93 | 340.6 | 1.002 | 0.031 | 61.00 | 0.39 |
| WCYN | 34.0844 | -107.0837 | 7.4 | 283.1 | 103.0 | 21.81 | 341.1 | 0.605 | 0.034 | 125.18 | 8.03 |

**Table S3:** Station-level measurements for the October 16, 2024 (BB2) 10-ton surface explosion. Columns: station ID; latitude, longitude; geodesic range (km) from 34.0693° N, 107.00497° W; geometric azimuth and back azimuth (degrees from north); signal travel time relative to the detonation; apparent celerity; dominant signal period (s) with uncertainty; peak-to-peak amplitude (Pa) with uncertainty. Signal period and amplitude are means over all band sets (see Section 2.2 in the main text). All stations are single sensors; azimuths are geometric, not direction-of-arrival estimates.

| BB2 Station code | Latitude | Longitude | Distance [km] | Azimuth [deg] | Back azimuth [deg] | Observed travel time [s] | Celerity [m/s] | Dominant signal period | Error | Peak-to-peak amplitude [Pa] | Error |
|---|---|---|---|---|---|---|---|---|---|---|---|
| ANTO | 33.8987 | -106.8782 | 22.28 | 148.3 | 328.4 | 65.40 | 340.7 | 0.907 | 0.069 | 10.73 | 0.03 |
| BXCYN | 34.0112 | -106.9929 | 6.56 | 170.2 | 350.2 | 19.17 | 342.0 | 0.609 | 0.004 | 89.70 | 2.64 |
| DUNT | 34.1627 | -106.9014 | 14.10 | 42.5 | 222.6 | 40.52 | 348.0 | 0.625 | 0.010 | 71.18 | 4.19 |
| ESDA | 34.1017 | -106.8975 | 10.53 | 70.0 | 250.0 | 32.55 | 323.5 | 0.607 | 0.013 | 74.30 | 1.47 |
| GRDY | 34.1458 | -106.8031 | 20.44 | 65.4 | 245.5 | 59.14 | 345.7 | 0.268 | 0.024 | 47.13 | 7.27 |
| MBP01 | 34.0804 | -106.9354 | 6.53 | 79.0 | 259.1 | 19.16 | 340.6 | 0.508 | 0.012 | 173.99 | 6.20 |
| MBP02 | 34.0813 | -106.9252 | 7.47 | 79.7 | 259.7 | 23.92 | 312.1 | 0.543 | 0.016 | 122.24 | 3.85 |
| MBP03 | 34.0813 | -106.9252 | 7.47 | 79.7 | 259.8 | 23.05 | 323.9 | 0.612 | 0.015 | 120.05 | 1.30 |
| MMTN | 34.0839 | -107.0576 | 5.11 | 288.5 | 108.5 | 14.54 | 351.5 | 0.426 | 0.017 | 221.71 | 10.34 |
| MZA01 | 34.1357 | -106.9835 | 7.64 | 15.0 | 195.0 | 21.70 | 352.1 | 0.500 | 0.010 | 137.48 | 4.09 |
| MZA02 | 34.1168 | -106.9296 | 8.72 | 52.7 | 232.8 | 25.07 | 347.9 | 0.444 | 0.013 | 123.43 | 8.54 |
| MZA03 | 34.0290 | -106.932 | 8.08 | 123.7 | 303.7 | 23.67 | 341.3 | 0.598 | 0.011 | 78.44 | 1.21 |
| MZA05 | 34.0303 | -107.0343 | 5.11 | 211.9 | 31.9 | 14.92 | 342.6 | 0.532 | 0.014 | 152.66 | 3.20 |
| MZA07 | 34.1077 | -107.0369 | 5.18 | 325.4 | 145.4 | 14.65 | 353.7 | 0.414 | 0.015 | 211.86 | 10.75 |
| NRMN | 34.0620 | -106.9117 | 8.63 | 95.4 | 275.4 | 25.25 | 341.9 | 0.535 | 0.006 | 108.55 | 5.17 |
| PICO | 34.0685 | -106.9682 | 3.39 | 91.4 | 271.5 | 9.75 | 347.6 | 0.463 | 0.021 | 251.12 | 8.77 |
| PPL01 | 34.0880 | -106.9855 | 2.74 | 40.8 | 220.8 | 7.66 | 357.9 | 0.394 | 0.012 | 252.46 | 10.65 |
| PPL05 | 34.1587 | -106.9588 | 10.81 | 23.1 | 203.1 | 30.89 | 350.1 | 0.536 | 0.010 | 102.12 | 6.24 |
| PPL06 | 34.1758 | -106.9519 | 12.81 | 22.4 | 202.4 | 36.62 | 349.9 | 0.584 | 0.010 | 60.65 | 3.71 |
| PPL07 | 34.1933 | -106.9446 | 14.87 | 21.9 | 202.0 | 42.54 | 349.6 | 0.589 | 0.025 | 70.40 | 4.79 |
| PPL10 | 34.2452 | -106.9203 | 21.05 | 21.7 | 201.7 | 60.35 | 348.9 | 0.287 | 0.013 | 95.99 | 11.07 |
| SLVR | 34.2563 | -106.9169 | 22.32 | 21.3 | 201.3 | 63.97 | 348.8 | 0.334 | 0.010 | 91.68 | 7.32 |
| WCL01 | 34.0598 | -107.0424 | 3.60 | 253.0 | 72.9 | 10.33 | 348.5 | 0.461 | 0.016 | 227.01 | 5.50 |
| WCL02 | 34.0579 | -107.0496 | 4.30 | 252.8 | 72.8 | 12.37 | 347.9 | 0.455 | 0.011 | 181.67 | 6.41 |
| WCL03 | 34.0559 | -107.0571 | 5.02 | 252.8 | 72.8 | 16.48 | 304.9 | 0.471 | 0.011 | 154.35 | 6.25 |
| WCL04 | 34.0539 | -107.0643 | 5.72 | 252.6 | 72.6 | 16.52 | 346.4 | 0.473 | 0.016 | 126.27 | 5.43 |
| WCL06 | 34.0501 | -107.0787 | 7.12 | 252.6 | 72.5 | 20.59 | 345.8 | 0.483 | 0.014 | 73.56 | 2.64 |
| WCL07 | 34.0478 | -107.0857 | 7.81 | 252.2 | 72.1 | 22.61 | 345.6 | 0.501 | 0.012 | 53.41 | 1.88 |
| WCL11 | 34.0406 | -107.12 | 11.06 | 253.3 | 73.2 | 32.07 | 345.0 | 0.670 | 0.011 | 16.45 | 0.20 |
| WCNRD | 34.0618 | -107.1006 | 8.85 | 264.6 | 84.6 | 25.51 | 347.0 | 0.565 | 0.013 | 35.97 | 3.53 |
| WCYN | 34.0844 | -107.0837 | 7.44 | 283.1 | 103.0 | 21.29 | 349.5 | 0.387 | 0.021 | 119.37 | 13.08 |

## S.1 Statistical Significance

Formal statistical tests (Tables S4 and S5) confirm the significance of the observations described above. The contrasting seasonal states are therefore not subtle fluctuations but significant, repeatable differences reflecting fundamental changes in tropospheric structure.

To quantify the bifurcation observed in May, the network was divided into two azimuthal groups: PPL+SLVR stations (azimuth ~22°) and WCL stations (azimuth ~253°). This configuration was explicitly designed to test for anisotropy. Welch's *t*-tests (Ruxton, 2006) were applied to both the full dataset and a distance-limited subset (stations ≤15 km) with a significance threshold of $\alpha = 0.05$.

The May 2024 (BB1) records reveal a statistically robust celerity divergence between azimuthal groups. Across all stations, the 22° azimuth line exhibits a mean apparent celerity of 346.9 m/s compared to 342.8 m/s for the 253° azimuth line, a difference significant at $p < 0.001$ (Table S4). Mean signal periods, however, are indistinguishable at this scale (0.76 s versus 0.78 s; $p = 0.81$; Table S5). When the analysis is restricted to stations within 15 km, the anisotropy becomes more distinct: celerities remain significantly different ($p = 0.0017$), and the signal periods diverge as well, with PPL+SLVR stations averaging 0.57 s and WCL stations averaging 0.78 s ($p = 0.0139$). These results demonstrate that the physical mechanism responsible for period differentiation is most strongly expressed in the proximal wavefield, while range-dependent processes such as attenuation, multipathing, or topographic interactions may obscure it at greater distances. Future deployments that extend the azimuthally aligned network to larger ranges will be essential for testing this interpretation.

In contrast, the October 2024 (BB2) experiment shows no significant evidence of directional (azimuthal) dependence (Figures 2, 4 in the main text). Neither celerity nor period differs significantly between station groups, whether all ranges are considered or the analysis is confined to within 15 km (all $p > 0.1$). This outcome confirms that the anisotropy observed in May reflects a seasonally contingent atmospheric state rather than a persistent geometric artifact. These statistical results provide quantitative confirmation of the anisotropy and motivate a broader consideration of its implications for monitoring and source characterization.

**Table S4:** Summary statistics for apparent celerity measured during the May (BB1) and October (BB2) 2024 campaigns. Values include the number of stations (*n*), mean and standard deviation of celerity, Welch's t-test results comparing azimuthal groups, and linear regression (LR) parameters of celerity versus distance (slope, intercept, and coefficient of determination $R^2$). Significant differences ($p < 0.05$) are observed between azimuthal groups in May, particularly for stations within 15 km, while no significant anisotropy is detected in October.

| Season (Campaign) | Dataset | Stations (n) | Celerity, average [m/s] | Celerity, standard deviation [m/s] | Welch's t-test (p) | LR slope | LR intercept | $R^2$ |
|---|---|---|---|---|---|---|---|---|
| May 2024 (BB1) | PPL+SLVR (all) | 8 | 346.89 | 1.88 | <0.001 vs WCL | -0.2862 | 351.0393 | 0.89 |
| | WCL (all) | 9 | 342.83 | 1.67 | — | -0.8228 | 348.1049 | 0.88 |
| May 2024 (BB1) | PPL+SLVR (≤15 km) | 4 | 348.29 | 1.66 | 0.0017 vs WCL | -0.4283 | 352.2843 | 0.76 |
| | WCL (≤15 km) | 9 | 342.83 | 1.67 | — | -0.8228 | 348.1049 | 0.88 |
| Oct 2024 (BB2) | PPL+SLVR (all) | 8 | 349.46 | 0.56 | >0.1 vs WCL | -0.1111 | 351.2749 | 1.00 |
| | WCL (all) | 9 | 340.59 | 15.79 | — | 1.0058 | 334.1726 | 0.03 |

**Table S5:** Summary statistics for signal period measured during the May (BB1) and October (BB2) 2024 campaigns. Values include the number of stations (*n*), mean and standard deviation of period, Welch's t-test results comparing azimuthal groups, and linear regression (LR) parameters of period versus distance (slope, intercept, and coefficient of determination $R^2$). Period differences become statistically significant for the May dataset within 15 km, whereas no significant anisotropy is observed in October.

| Season (Campaign) | Dataset | Stations (n) | Period, average [s] | Period, standard deviation [m/s] | Welch's t-test (p) | LR slope | LR intercept | $R^2$ |
|---|---|---|---|---|---|---|---|---|
| May 2024 (BB1) | PPL+SLVR (all) | 8 | 0.758 | 0.208 | 0.8069 vs WCL | 0.0323 | 0.2891 | 0.92 |
| | WCL (all) | 9 | 0.781 | 0.168 | — | 0.0827 | 0.2507 | 0.88 |
| May 2024 (BB1) | PPL+SLVR (≤15 km) | 4 | 0.573 | 0.085 | 0.0139 vs WCL | 0.0226 | 0.3629 | 0.81 |
| | WCL (≤15 km) | 9 | 0.780 | 0.168 | — | 0.0827 | 0.2507 | 0.88 |
| Oct 2024 (BB2) | PPL+SLVR (all) | 8 | 0.511 | 0.120 | >0.1 vs WCL | -0.0207 | 0.8257 | 0.75 |
| | WCL (all) | 9 | 0.502 | 0.076 | — | 0.0269 | 0.3307 | 0.82 |